\documentclass[preprint,showpacs,aps]{revtex4}
\usepackage[dvips]{epsfig}
\newcommand{\clW}{{\cal W}}

\newcommand{\prt}{\partial}

\newcommand{\rgl}{\rangle}
\newcommand{\lgl}{\langle}

\begin{document}

\title{The Localization Length of Stationary States in the Nonlinear
Schr\"odinger Equation}

\author{Alexander Iomin and Shmuel Fishman}

\affiliation{Department of Physics, Technion, Haifa, 32000,
Israel}

\date{\today}
\begin{abstract}
For the nonlinear Schr\"odinger equation (NLSE), in presence of disorder,
exponentially localized stationary states are found. In the present work 
it is demonstrated analytically that the localization length is typically 
independent of the strength of the nonlinearity and is identical  to the 
one found for the corresponding linear equation. The analysis makes use of 
the correspondence between the stationary NLSE  and the Langevin equation 
as well as of the resulting Fokker-Planck equation. The calculations are 
performed for the ``white noise'' random potential and an exact expression 
for the exponential growth of the eigenstates is obtained analytically. 
It is argued that the main conclusions are robust.

\end{abstract}


\maketitle

In this work we consider a simple problem of the one-dimensional
Anderson localization \cite{Anderson,Lee} for the nonlinear Schr\"odinger 
equation (NLSE) \cite{bwFSW,bwDS,kivshar}. This problem is relevant to 
experiments in nonlinear optics, for example disordered
photonic lattices \cite{f1}, where Anderson localization was found
in presence of nonlinear effects as well as 
experiments on Bose-Einstein Condensates (BEC) in disordered
optical lattices \cite{BECE1,BECE3,akkermans,SanchAspect,BShapiro}. 
The interplay between disorder and nonlinear effects leads to new
interesting physics \cite{BECE3,akkermans,bishop1,aubry,mackay,aub1}.
In particular, the problem of spreading of wave packets and transmission 
are 
not simply related \cite{doucot, pavloff}, in contrast with the linear 
case.
In spite of the extensive research, many fundamental problems are still 
open, and, in particular, it is not clear whether in one 
dimension (1D) Anderson localization can survive the effects of 
nonlinearities.

Herein we consider 1D localization of stationary solutions of the
NLSE in a random potential. The problem is described by the
equation
\begin{equation}\label{nlse1}
i\prt_t\psi=-\prt_{x}^2\psi+\beta|\psi|^2\psi+V(x)\psi\, ,
\end{equation}
where $V(x)$ is a random $\delta$-correlated potential with a
Gaussian distribution, of zero mean and variance $\sigma^2$, such
that
\begin{equation}\label{nlse2}
\lgl V(x)V(x')\rgl=2\sigma^2\delta(x-x')\, ,
\end{equation}
where $ \lgl\dots\rgl$ denotes the average over realizations of the random 
potential.
The variables are chosen in dimensionless units and the Planck
constant is $\hbar=1$. For the linear case $(\beta=0)$ this model was 
studied extensively in the past \cite{halperin, LGP}.
The problem in question is Anderson
localization of stationary solutions of Eq. (\ref{nlse1}) with
energies $\omega$
\begin{equation}\label{nlse3}
\psi(x,t)=\exp(-i\omega t) \phi(x)\, ,
\end{equation}
where $\phi(x)$ is real. Substituting Eq. (\ref{nlse3}) in Eq.
(\ref{nlse1}) 
one obtains the stationary NLSE 
\begin{equation}\label{nlse_A1}
\omega\phi(x)=-\prt_{x}^2\phi(x)+\beta\phi^3(x)+V(x)\phi(x)\, .
\end{equation}
It was established rigorously \cite{bwFSW,AF1,AF2} that this equation has 
exponentially localized solutions of this type for a wide range 
conditions. 
It is instructive to notice that also in absence of disorder $(V=0)$ the 
nonlinear equation exhibits stationary localized states, and these are not
simply related to the localized states in presence of disorder
\cite{flach,mackay,aubry}. The purpose 
of the present work is to find the localization length of these states.
It will turn out that the localization length is typically not affected by 
the 
nonlinearity and is {\it identical} to the one of the linear problem
$(\beta=0)$. 
The approach {\it \'a la} Borland \cite{borland,mott} will be the basis of 
our analysis. This approach is reviewed clearly in detail in 
\cite{delyonA} and made rigorous in \cite{delyonB}.

We will specifically calculate $\lgl\phi^2(x)\rgl$ of 
solutions of Eq.
(\ref{nlse_A1}) that are found for a certain $\omega$, with given boundary 
conditions at some point, for example $\phi(x=0)$ and 
$\phi^{\prime}(x=0)$,
where prime means the derivative with respect to $x$. It will be shown 
that this quantity grows exponentially with the rate 
\begin{equation}\label{nlse_A2}
2\gamma=\lim_{x\to\infty}\frac{ \ln\lgl\phi^2(x)\rgl}{x}>0\, , 
~~~\xi=\frac{1}{\gamma}\, ,
\end{equation}
that is independent of $\beta$, where $\xi$ is the localization length. 
Note, it is different from the usually studied self averaging quantity 
$\gamma_s=\frac{1}{2}\frac{d}{dx}\lgl\ln\phi^2(x)\rgl=
\frac{1}{2}\lim_{x\to\infty}\frac{\ln\phi^2(x)}{x}$. 
We will find that $\gamma$ is 
a smooth function of energy. Since the distribution of random potentials
is translationally invariant, it is independent of the choice of the 
initial point as $x=0$. Like in the linear case, starting from a specific 
initial condition, $\phi(x)$ will typically grow. For specific values of 
$\omega$ 
at some point this function will start to decay, so that a normalized
eigenfunction is found \cite{borland,mott,delyonA,delyonB}. 
The envelope of the wave function will grow exponentially if we start 
either from the right or from the left. The value of $\omega$ results from 
the matching condition, so that an eigenfunction has some maximum and 
decays in both directions as required by the normalization condition. The 
exponential decay is an asymptotic property, while the matching is 
determined by the potential in the vicinity of the maximum. This 
observation \cite{delyonA,delyonB} is crucial for the validity of this 
approach and 
enables us to determine the exponential decay rate of states from the 
solution of the initial value problem (\ref{nlse_A1}). For the 
linear case these values of 
$\omega$ form the point spectrum of the problem that is the entire 
spectrum of the linear problem. This approach can be followed also 
for the nonlinear problem, but contrary to the linear case, these 
stationary states do not provide a complete picture of the dynamics.
Let us fix $\omega$, the stationary states in the vicinity of $\omega$ 
will on the average (over realizations) decay with the localization length 
$\xi$ defined by Eq. (\ref{nlse_A2}). As mentioned it will be found to be 
independent of $\beta$, leading to the conclusion that this is a typical 
property of the localized eigenstates of Eq. (\ref{nlse_A1}).

The calculation of $\lgl\phi^2(x)\rgl$ will be performed by the analogy 
with the classical Langevin equation \cite{halperin,LGP}. 
Therefore, here we are considering the $x$-coordinate as the formal
time on the half axis $x\equiv\tau\in[0,\infty)$, and Eq. (\ref{nlse_A1})
reduces to the Langevin equation
\begin{equation}\label{nlse4}
\ddot{\phi}+\omega\phi-\beta\phi^3-V(\tau)\phi=0
\end{equation}
with the $\delta$ correlated Gaussian noise $V(\tau)$.
Now we
introduce new variables $u=\phi$ and $v=\dot{\phi}\equiv 
\frac{d\phi}{d\tau}$ and a
distribution function of these new variables is $P=P(u,v,\tau)$. 
The dynamical process in the presence of the Gaussian
$\delta$-correlated noise is described by the distribution
function that satisfies the Fokker-Planck equation:
(FPE) (see, \textit{e.g.}, \cite{risken,rytov})
\begin{equation}\label{nlse5}
\prt_{\tau}P-[\omega u-\beta 
u^3]\prt_vP+v\prt_uP-\sigma^2 u^2\prt_{v}^2P=0\, .
\end{equation}
It is obtained from the Langevin equation (\ref{nlse4}), as can be seen 
in \cite{risken,rytov} and is consistent with \cite{LGP}, Eq. (6.14) there
\cite{add1}.

We are interested in the average quantum probability density
$\lgl\phi^2(x)\rgl\equiv \lgl u^2(\tau)\rgl$, where
\[\lgl u^2(\tau)\rgl=\int u^2P(u,v,\tau)dudv\, .\]
It is useful to obtain from the FPE a system of equations for the
moments
\begin{equation}\label{nlse6}
M_{k,l}=\lgl u^kv^l\rgl\, ,
\end{equation}
where $k,l=0,1,2,\dots$.
Substituting $u^kv^l$ in the FPE and integrating over $u$ and $v$,
one obtains the following relation for $M_{k,l}$
\begin{equation}\label{nlse7}
\dot{M}_{k,l}=-l\omega M_{k+1,l-1}+kM_{k-1,l+1} 
+l(l-1)\sigma^2 M_{k+2,l-2} +\beta l M_{k+3,l-1}\, ,
\end{equation}
where $ M_{k,l}$ with negative indexes are assumed to vanish.
We note that only terms with the same parity of $k+l$ are coupled. Since 
we are interested in $M_{2,0}=\lgl u^2\rgl$, we study only the case when 
this parity is even, namely $k+l=2n$ with $n=1,2,\dots$.
  The sum of the indexes of the moments
is $2n$, except the last term $\beta l M_{k+3,l-1}$, where the sum
is $2(n+1)$. This leads to the infinite system of linear equations
that can be written in the form
\begin{equation}\label{nlse8}
\dot{\bf M}=\clW{\bf M}\, ,
\end{equation}
where ${\bf M}=\left(M_{2,0},M_{1,1},M_{0,2},M_{4,0},M_{3,1},\dots\right)$
and $=\clW$ is the corresponding matrix.
The matrix elements $\clW_{k,l}$ are determined by Eq.
(\ref{nlse7}). The solutions of  the system of linear 
equations (\ref{nlse8}) are linear combinations of the eigenfunctions
\begin{equation}\label{nlse9}
M_{\lambda}(t)=\exp(\lambda\tau)M_{\lambda}(0)\, ,
\end{equation}
where $M_{\lambda}(0)$ is the eigenvector of $\clW$ corresponding to 
$\lambda$. The growth rate of each moment, in particular 
$M_{2,0}=\langle u^2\rangle$, is determined by the eigenvalue with the 
largest 
real part ${\rm Re\,}\lambda$.

For $\beta=0$, Eq. (\ref{nlse7}) (or (\ref{nlse8})) has a closed form for 
each $n$.
Therefore, the infinite matrix $\clW$ is block diagonal and consists of 
the independent blocks $A_n = A_n[(2n+1)\times (2n+1)]$, and the
characteristic polynomial reduces to a product of their determinants
\begin{equation}\label{nlse10}
\prod_{n=1}^{\infty}{\rm det}\left(A_n-\lambda I_n\right)=0\, ,
\end{equation}
where $I_n$ is an $(2n+1)\times(2n+1)$ unit matrix. The problem of
localization in the framework of Eq. (\ref{nlse7}) reduces to $n=1$. The 
relevant characteristic polynomial ${\rm det\,}(A_1-\lambda I_1)$ reduces 
to a cubic equation
\begin{equation}\label{nlse11}
\lambda^3+4\omega\lambda-4\sigma^2=0\, .
\end{equation}
Cardano's method yields \cite{korn}
\begin{equation}\label{nlse12}
\lambda_1=R_{+}+R_{-}\, ,
~~~\lambda_{2,3}=-\frac{R_{+}+R_{-}}{2}\pm
i\sqrt{3}\frac{R_{+}-R_{-}}{2}\, ,
\end{equation}
where $R_{\pm}=\left[2\sigma^2\pm\sqrt{4\sigma^4+
\frac{64\omega^3}{27}}\right]^{\frac{1}{3}}$. 
The growth rate is determined by the eigenvalue with largest real part
that will be denoted by $\lambda_m$. We conclude that asymptotically 
for large $x$ the averaged wave 
function indeed grows exponentially as 
$\lgl\phi^2(x)\rgl\sim e^{2\gamma x}$ and $2\gamma=\lambda_m$.
Also the behavior of the higher blocks can be calculated. From the $n$-th 
block the behavior of the moment $\lgl\phi^{2n}(x)\rgl$ can be found. In 
the high energy limit one finds from Eq. (\ref{nlse12})
\begin{equation}\label{lgp_1}
\lambda_m=\lambda_1\approx \frac{\sigma^2}{\omega}\, ,
~~~~\omega\rightarrow +\infty\, ,
\end{equation}
\begin{equation}\label{lgp_2}
\lambda_m={\rm Re\,}\lambda_3\approx 2\sqrt{|\omega|}\, , ~~~~
\omega\rightarrow -\infty\, .
\end{equation}
These limits can be found directly from Eq. (\ref{nlse11}).
The solutions should be compared with the high energy asymptotics obtained 
in Ref \cite{LGP} (Eq. (10.12), p. 143), where 
$\gamma_s=\sigma^2/4\omega$, and $\gamma_s=\sqrt{|\omega|}$ in limits 
$\omega\rightarrow +\infty$ and $\omega\rightarrow -\infty$, respectively.
  Note that Eq. (\ref{nlse12}) gives a 
simple expression for $\gamma$ for {\it all} values of the parameters, 
while the expression for $\gamma_s$ is known only in the large $\omega$ 
limit. Since $\gamma$ and $\gamma_s$ result of different averages, these 
are not expected to be identical.

Now, let us consider localization for $\beta\neq 0$. The
eigenvalue equation 
\begin{equation}\label{nlse13}
\clW(\beta)M_{\lambda}(\beta)=\lambda(\beta)M_{\lambda}(\beta)\, ,
\end{equation}
is obtained from Eq. (\ref{nlse7}), $\clW$ is not block diagonal 
anymore. The $\beta$ dependence results from the last term in Eq. 
(\ref{nlse7}). The $\beta$ dependent terms couple the $n$-th block
and $(n+1)$ block and are located above the $(n+1)$ block and to the right 
of the $n$-th block. Consequently, the $\beta$ dependent terms do not 
affect the characteristic polynomial, as can be shown by elementary 
operations on determinants. It reduces to the one found for $\beta=0$,
namely Eq. (\ref{nlse10}). Therefore, growth rates of all moments of 
$\phi(x)$ do not depend on $\beta$ and their values are equal to the ones 
of the linear problem for $\beta=0$. This is correct in particular for 
$\lambda_m$, consequently $\gamma$ of Eq. (\ref{nlse_A2}) is identical
to the value found in the linear case $(\beta=0)$.

We demonstrated that in the presence of a random potential the stationary 
states of the NLSE are exponentially localized with the {\it localization 
length that is found in the absence of the nonlinearity}. This is in 
agreement with a heuristic argument that the effect of nonlinearity is 
negligible where the wave function is small. We believe that the approach 
{\it \'a la} Borland can be extended to a rigorous treatment of the 
stationary states of the NLSE.
Since this equation is nonlinear, the stationary states do not provide the 
complete or even essential description of the dynamics, starting 
from a given initial condition. The status of stability of these states 
with respect of small perturbations is not clear. 
The relation to the transmission problem is not obvious. Our results are 
consistent with the limit of vanishing flux in the transmission problem 
\cite{PavloffLeboeuf}.

Another question that should be discussed is of the generality of the 
results.  Assume that the power of $\psi$ in Eq. (\ref{nlse1}) differs 
from $3$. Only the last term in Eq. (\ref{nlse7}) will be affected 
resulting in a different coupling between the blocks of the matrix 
$\clW$. But since 
these couplings are above the diagonal of the block diagonal matrix,
they will not affect our conclusion that the characteristic 
polynomial which determines $\lambda$ is not affected by the nonlinearity.
If the potential $V(x)$ deviates from a white noise one, Kramers-Moyal 
coefficients \cite{risken} that are higher than the second one appear. But 
if no convergence problems of the Kramers-Moyal expansion are encountered,
an equation like Eq. (\ref{nlse5}) with higher powers of $u$ and $v$, 
combined with higher order derivatives is obtained. Because of the 
structure of the Kramers-Moyal expansion the block diagonal form of the 
matrix 
$\clW$ in absence of the nonlinearity is expected to be unaffected by this 
deviation from white noise. Therefore we expect the main result of the 
work, namely the independence of the localization length of the 
nonlinearity to exhibit some degree of robustness and to hold for a wide 
range of models, beyond 
the specific model that is studied in detail in the present work. All
these problems should be subject of further studies.

This work was supported in part by the Israel Science Foundation
(ISF), by the US-Israel Binational Science Foundation (BSF), and
by the Minerva Center for Nonlinear Physics of Complex Systems. It
is our great please to thank I. Guarneri and A. Soffer for very critical 
discussions that affected some directions of our work and 
E. Baskin, S. Flach, Y. Krivolapov, B. Shapiro, and D. Shepelyansky
for very informative and instructive discussions and communications.
We thank S. Aubry, S. Gredeskul, P. Leboeuf, and N. Pavloff for 
communications and comments after the work was completed.


\begin{thebibliography}{10}
\bibitem{add1} To see this one has to define $P(z,\tau)=\int dudv 
\delta\left(z-\frac{v}{u}\right)P(u,v,\tau)$ and derive Eq. (6.14) of 
\cite{LGP} from Eq. (\ref{nlse5}).
\end{thebibliography}
\end{document}